\documentclass[prl,aps,twocolumn,10pt,showpacs,groupedaddress,superscriptaddress,floatfix]{revtex4-1}
\usepackage{amsmath,amsfonts,amssymb,graphics,graphicx,epsfig,color,times}%,bbm}
\usepackage{color}
\usepackage{hyperref}
\usepackage{mathrsfs}
\usepackage{verbatim}
\begin{document}
\title{Spectral Shaping of Cascade Emissions from Multiplexed Cold Atomic Ensembles}

\author{H. H. Jen}
\email{sappyjen@gmail.com}
\affiliation{Institute of Physics, Academia Sinica, Taipei 11529, Taiwan}
\author{Y.-C. Chen}
\affiliation{Institute of Atomic and Molecular Sciences, Academia Sinica, Taipei 10617, Taiwan}
\date{\today}
\renewcommand{\k}{\mathbf{k}}
\renewcommand{\r}{\mathbf{r}}
\newcommand{\f}{\mathbf{f}}

\begin{abstract}
We investigate the spectral properties of the biphoton state from the cascade emissions of cold atomic ensembles, which composes of a telecommunication photon (signal) followed by an infrared one (idler) via four-wave mixing.\ With adiabatic conditions for Gaussian driving pulses of width $\tau$, the spectrum of the biphoton state has the form of a Gaussian that conserves signal and idler photon energies within $\hbar/\tau$ modulated by a Lorentzian with a superradiant linewidth.\ Multiplexing the atomic ensembles with frequency-shifted cascade emissions, we may manipulate and shape the spectrum of the biphoton state.\ The entropy of entanglement is derived from Schmidt decomposition, which can be larger if we multiplex the atomic ensembles in a way that conserves signal and idler photon central energies.\ The eigenvalues in Schmidt bases are degenerate in pairs for symmetric spectral shaping in which the mode probability densities show interference patterns.\ We also demonstrate the excess entropy of entanglement that comes from continuous frequency space, which scales up the total entropy.\ The scheme of multiplexed cascade-emitted biphoton state provides multimode structures that are useful in long-distance quantum communication and multimode quantum information processing.
\end{abstract}
\maketitle
Long-distance quantum communication is a challenging task for requirements of stability and coherence in the information carriers.\ To overcome this difficulty, a quantum repeater protocol \cite{Briegel1998, Dur1999} enables the distribution of quantum information by inserting quantum memory elements between distant information receivers before coherence deteriorates.\ Quantum memories using the setup of cold atomic ensembles (AE) in long-distance quantum communication \cite{Duan2001} are advantageous for its controllability and efficient manipulation of Raman-type light-matter interactions \cite{Matsukevich2004,Chou2004,Chaneliere2005, Chen2006, Laurat2006}.\ Furthermore the telecommunication (telecom) bandwidth in the cascade emissions \cite{Chaneliere2006} is beneficial for low-loss optical fiber transmission, and it can even be frequency-converted to infrared wavelength for storage \cite{Radnaev2010, Jen2010}.

Quantum communication carriers and quantum storage via light-matter interaction are not limited to discrete degrees of freedom, e.g. polarizations \cite{Clauser1969, Aspect1981, Kwiat1995} or frequencies \cite{Lan2007,Ramelow2009} of light.\ The continuous entanglement can provide unlimited communication capacity in the transverse momentum \cite{Law2004, Moreau2014}, space \cite{Grad2012}, and energy-time domains \cite{Branning1999, Law2000, Parker2000}.\ This opens up a greater potentiality in quantum cryptography \cite{Gisin2002} and quantum information applications \cite{Braunstein2005}.\ Other useful degrees of freedom involve orbital angular momenta of light \cite{Arnaut2000, Mair2001, Molina2007} which recently can create entanglement in high dimensions \cite{Dada2011, Agnew2011, Fickler2012} for quantum storage in a crystal \cite{Zhou2015} or atomic ensemble \cite{Nicolas2014, Ding2015}.\ Similar high dimensional quantum memories are proposed using atomic \cite{Afzelius2009} or optical \cite{Zheng2015} frequency combs, and moreover a speedup in quantum repeater protocol can be realized by multiplexing multimode quantum memories in space \cite{Collins2007, Lan2009} or time \cite{Simon2007}.\ Recently spectral shaping \cite{Bernhard2013, Lukens2014} in the biphoton state generated from spontaneous parametric down conversion provides methods to manipulate the spectral information with full control, en route to constructing multimode quantum communication.

For the system of cold AE as an alternative platform of quantum information processing, efficiently controlled and low-loss long-distance quantum communication in continuous frequency space has not been investigated.\ In this Letter we propose spectral shaping of the cascade emissions from cold AE, which are frequency multiplexed by frequency shifters.\ We first formulate the spectra of the biphoton state, and analyze the multimode structures via Schmidt decomposition.\ The entanglement of such scheme can be controlled by independently shifting the central frequencies of the cascade emissions, and we show pairwise degeneracies in the mode probability densities which indicate interference patterns.\ We also identify the excess entropy from the frequency entanglement for our proposed multiplexed scheme, which scales up the total Hilbert space exponentially. 
%%%%%%%%%%%%%%%%%%%%%%%%%%%%%%%%%%%%%%%%%%%%%%%%%%%%%%%%%%%%%%%%%%%%%%%%%%%%%%%%%%%%%%%%%%%%%%%%%%%%%%%%%%%%%%%%%%%%

\begin{figure}[t]
\centering
\includegraphics[width=8.5cm,height=6cm]{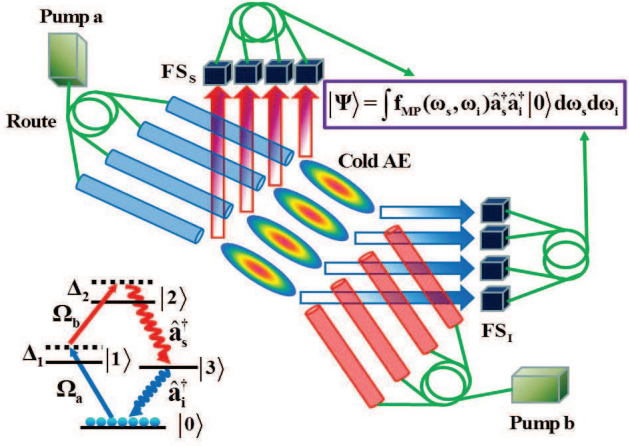}
\caption{(Color online) Schematic multiplexed cold atom ensembles (AE) and diamond-type atomic level.\ The signal and idler photon pair $\hat{a}^\dagger_{s,i}$ is generated by driving AE by two pump fields $\Omega_{a,b}$ with single and two-photon detunings $\Delta_{1,2}$.\ For illustration we plot four AE, and a circle represents a routing pathway for the pump fields (rods) and cascade-emitted photon pair (arrows).\ FS$_{\rm S(I)}$ stand for frequency shifters (e.g. acousto-optic modulators) of signal (S) and idler (I) photons respectively, which individually address the frequency shifts to the photon pair.\ $|\Psi\rangle$ is the effective multiplexed biphoton state with spectral distribution $f_{\rm MP}(\omega_s,\omega_i)$.}\label{fig1}
\end{figure}

We consider a Rb atomic ensemble with a diamond-type level structure as shown in Fig. \ref{fig1}.\ The cascade emissions are generated by driving the atomic system with two classical pulses.\ With dipole approximation of light-matter interactions \cite{QO:Scully}, the Hamiltonian can be written in the interaction picture as
\begin{eqnarray}
V_{\rm I}&=&-\sum_{m=1,2}\Delta_m\sum_{\mu=1}^N|m\rangle_\mu\langle m|-\sum_{m=a,b}\left(\frac{\Omega_m}{2}\hat{P}_m^\dag+{\rm h.c.}\right)\nonumber\\
&&-i\sum_{m=s,i}\bigg\{\sum_{\k_m,\lambda_m}g_m\hat{a}_{\k_m,\lambda_m}\hat{Q}_m^\dag e^{-i\Delta\omega_m t}-{\rm h.c.}\bigg\},
\end{eqnarray} 
where we let $\hbar$ $=$ $1$, and denote $\lambda_m$ as polarizations of photons.\ We define the collective dipole operators as $(\hat{P}_m^\dag,$ $\hat{Q}_m^\dag)$ $\equiv$ $\sum_\mu\hat{T}_\mu e^{i\k_T\cdot\r_\mu}$ where $\hat{T}_\mu$ denotes various dipole transition operators $|j\rangle_\mu\langle l|$ associated with spatial phases induced by $\k_T$.\ The Rabi frequencies of two driving pulses are $\Omega_{a(b)}$ with central frequencies $\omega_{a(b)}$ and wavevectors $\k_{a(b)}$ while signal and idler photons have coupling constants $g_{s(i)}$ with central frequencies $\omega_{s(i)}$ and wavevectors $\k_{s(i)}$.\ Note that here we absorb $(\epsilon_{\k_m,\lambda_m}\cdot\hat{d}_m^*)$ into $g_{s(i)}$ for concise expressions, where $\epsilon_{\k_m,\lambda_m}$ is the polarization direction of the quantized bosonic fields $\hat{a}_{\k_m,\lambda_m}$, and $\hat{d}_m$ is the unit direction of the dipole operators.\ The single and two-photon detunings are $\Delta_1$ $=$ $\omega_a$ $-$ $\omega_1$ and $\Delta_2$ $=$ $\omega_a$ $+$ $\omega_b$ $-$ $\omega_2$ respectively, and we define $\Delta\omega_s$ $\equiv$ $\omega_s$ $-$ $\omega_2$ $+$ $\omega_3$ $-$ $\Delta_2$ and $\Delta\omega_i$ $\equiv$ $\omega_i$ $-$ $\omega_3$ where atomic level energies are $\omega_{1,2,3}$.\ The upper level $|2\rangle$ can be chosen 6S$_{1/2}$, 7S$_{1/2}$, or 4D$_{3/2(5/2)}$ that the telecom wavelength resides between 1.3-1.5 $\mu$m \cite{Chaneliere2006}.
%%%%%%%%%%%%%%%%%%%%%%%%%%%%%%%%%%%%%%%%%%%%%%%%%%%%%%%%%%%%%
%%%%%%%%%%%%%%%%%%%%%%%%%%%%%%%%%%%%%%%%%%%%%%%%%%%%%%%%%%%%%

With large detuned excitation pulses, the singly-excited atomic system adiabatically follows the driving pulses and spontaneously emits a telecom photon (signal) and subsequently an infrared one (idler) to the ground state.\ Four-wave mixing condition assures the highly correlated photon pair generation, and additionally the idler photon is observed to be superradiant \cite{Chaneliere2006} due to induced dipole-dipole interactions as a signature of collective radiation \cite{Dicke1954, Lehmberg1970}.\ From Schr\"odinger equation of motion, we may derive the biphoton state probability $D_{si}$ as \cite{Jen2012-2, SM}
\begin{eqnarray}
D_{si}(\Delta\omega_s,\Delta\omega_i)=\frac{\tilde{\Omega}_a\tilde{\Omega}_b g_s^*g_i^*\sum_\mu e^{i\Delta\k\cdot\r_\mu}}{4\Delta_1\Delta_2\sqrt{2\pi}\tau}f(\omega_s,\omega_i),\label{Dsi}
\end{eqnarray}
which depends on the excitation pulse areas $\tilde{\Omega}_{a(b)}$ and four-wave mixing condition $\Delta\k$ $=$ $\k_a$ $+$ $\k_b$ $-$ $\k_s$ $-$ $\k_i$.\ We define the spectral function of the biphoton state as
\begin{eqnarray}
f(\omega_s,\omega_i)=\frac{e^{-(\Delta\omega_s+\Delta\omega_i)^2\tau^2/8}}{\frac{\Gamma_3^{\rm N}}{2}-i\Delta\omega_i},
\end{eqnarray}
where $\tau$ is the pulse width of the driving fields.\ $\Gamma_3^{\rm N}/\Gamma_3$ $=$ $(\rm{N}\bar{\mu}+1)$ is the superradiant decay rate proportional to the optical density $\rm{N}\bar{\mu}$ with respectively the number of particles N and geometrical constant $\bar{\mu}$ \cite{Lehmberg1970}, and $\Gamma_3$ is the intrinsic (single-particle) spontaneous decay rate.\ The spectrum of the biphoton state has the form of a Gaussian with a maximum along $\Delta\omega_s$ $=$ $-\Delta\omega_i$ indicating the conservation of the total energy while two photons are entangled within the spectral range of $1/\tau$.\ It is modulated by a Lorentzian with a superradiant linewidth, which generates a more entangled biphoton state in frequency space as the optical density of the atomic ensemble increases \cite{Jen2012-2}.

The information of spatial correlation in the biphoton state can be removed by coupling to a single-mode fiber.\ We here consider a continuous frequency space in the state vector of such photon pair $|\Psi\rangle$, and for some specific polarizations $\lambda_s$ and $\lambda_i$, we have 
\begin{align}
|\Psi\rangle=\mathcal{N}\int f(\omega_s,\omega_i)\hat{a}_{\lambda_s}^\dag(\omega_s)\hat{a}_{\lambda_i}^\dag(\omega_i)|0\rangle d\omega_s d\omega_i,
\end{align}
where $\mathcal{N}$ denotes a dimension regularization that makes sure of dimensionless spectral function in the above.\ The overall constant of the biphoton state [see Eq. (\ref{Dsi})] indicates the generation probability (which is made small) while the spectral property only depends on $f(\omega_s,\omega_i)$.
%%%%%%%%%%%%%%%%%%%%%%%%%%%%%%%%%%%%%%%%%%%%%%%%%%%%%%%%%%%%%
\begin{figure}[b]
\centering
\includegraphics[width=8.5cm,height=4.8cm]{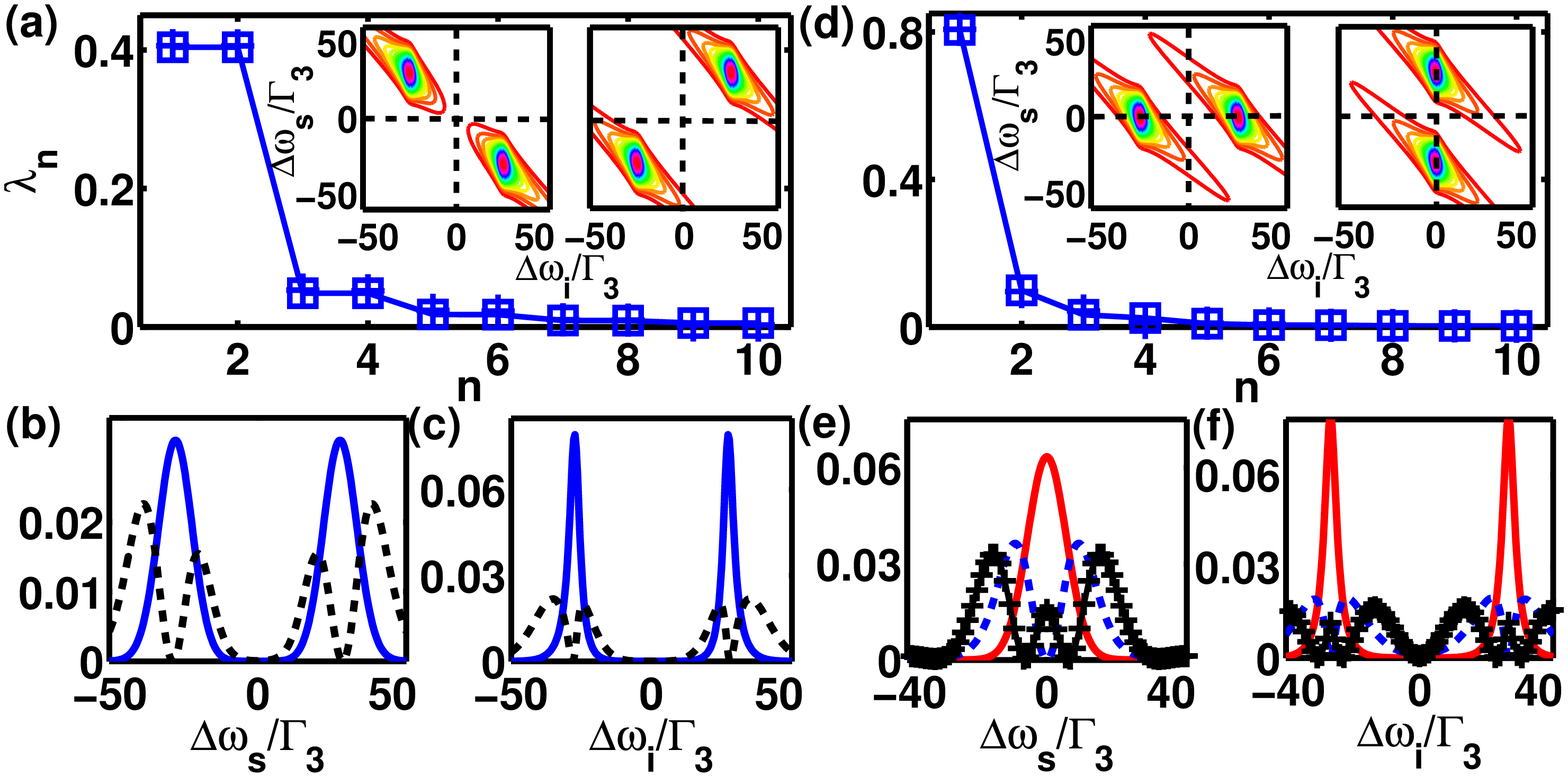}
\caption{(Color online) Spectral shaping in two cold AE.\ (a) Pairwise eigenvalues and absolute symmetric spectral distributions (insets) to the axis $\Delta\omega_s$ $+$ $\Delta\omega_i$ $=$ $0$, along with first four (b) signal $|\psi_n|^2$ and (c) idler mode probability densities $|\phi_n|^2$ (solid and dash for the first and the next two degenerate ones respectively) for the left inset of (a).\ (d) Eigenvalues and spectral distributions on $\Delta\omega_{s,i}$ $=$ $0$, and first three (e) signal and (f) idler mode probability densities (solid, dash, and $+$) for the left inset of (d).\ The eigenvalues are overlapped with each other ($\square$, $+$) for both insets respectively in (a) and (d) where dash lines guide the eyes for $\Delta\omega_{s,i}$ $=$ $0$.\ The spectral ranges for both signal and idler photons are post-selected to $\pm$ $300\Gamma_3$ throughout all figures where we also set $\Gamma^{\rm N}_3$ $=$ $5\Gamma_3$ and $\tau$ $=$ $0.25\Gamma_3^{-1}$ without loss of generality.\ Here we set $\delta p_1$ $=$ $30\Gamma_3$ in (a) and $\delta q_1$ $=$ $30\Gamma_3$ in (d).}\label{fig2}
\end{figure}
%%%%%%%%%%%%%%%%%%%%%%%%%%%%%%%%%%%%%%%%%%%%%%%%%%%%%%%%%%%%%

We can shape the spectral information of the cascade emissions by multiplexing multiple AE and shifting the frequencies of the signal and idler photons independently as shown in Fig. \ref{fig1}.\ Under the weak and common excitations as in the quantum repeater protocol of AE \cite{Duan2001}, up to the first nonvanishing order, we have the spectral function of the biphoton state in the form of
\begin{eqnarray}
f_{\rm MP}(\omega_s,\omega_i)=\sum_{m=1}^{\rm N_{\rm MP}}\frac{e^{-(\Delta\omega_s+\Delta\omega_i+\delta q_m)^2\tau^2/8}}{\frac{\Gamma_3^N}{2}-i(\Delta\omega_i+\delta p_m)},
\end{eqnarray}
where $\rm N_{\rm MP}$ is the number of multiplexed AE.\ $\delta p_m$ is the frequency shift from the frequency shifters of the idler photons only while $\delta q_m$ is the combined frequency shift from the frequency shifters of the signal and idler photons.\ Therefore these frequency shifts can be manipulated individually.
%%%%%%%%%%%%%%%%%%%%%%%%%%%%%%%%%%%%%%%%%%%%%%%%%%%%%%%%%%%%%
\begin{figure}[t]
\centering
\includegraphics[width=8.5cm,height=4.5cm]{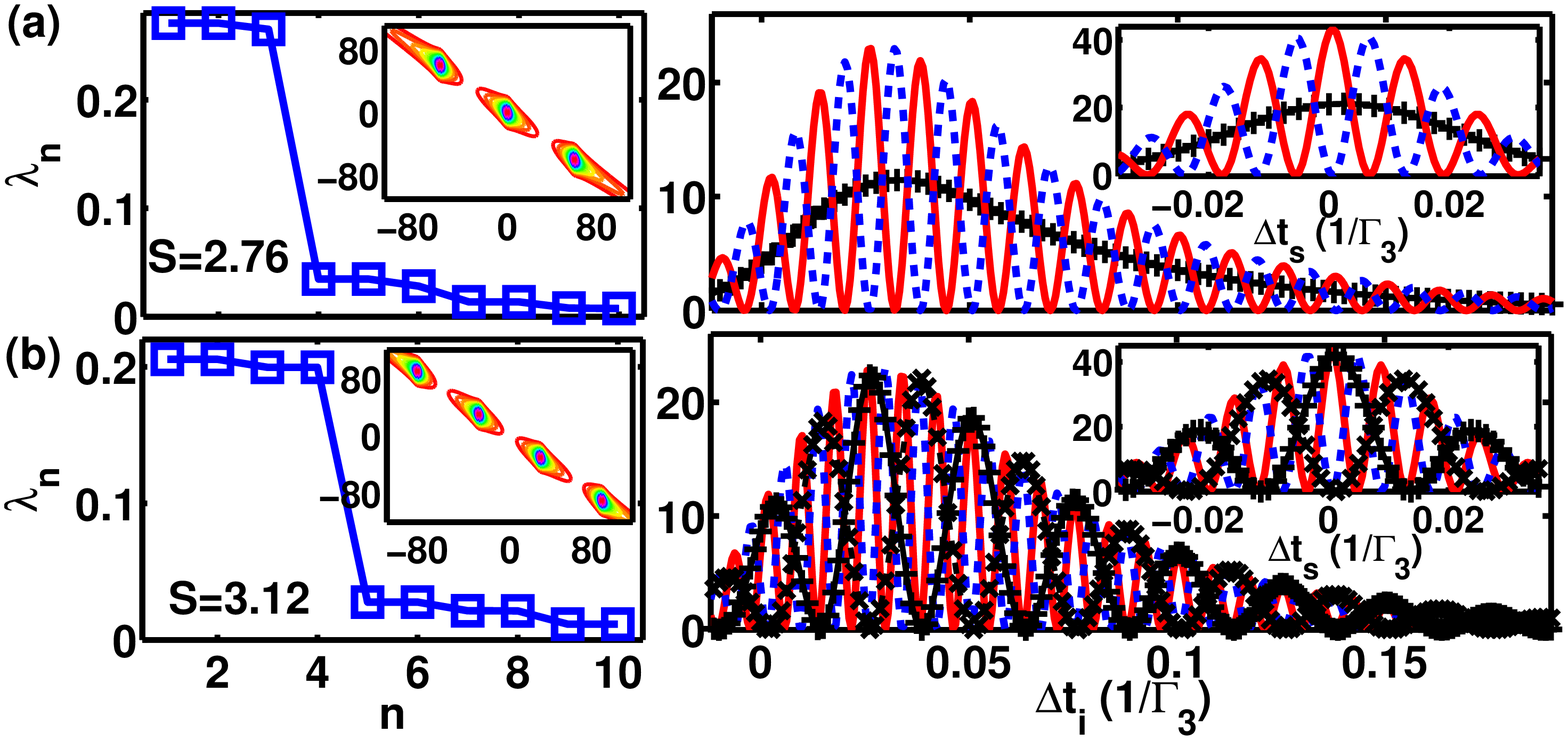}
\caption{(Color online) Symmetric spectral functions of three and four multiplexed cold AE and Fourier-transformed mode probability densities.\ The eigenvalues $\lambda_n$ ($\square$) in (a) and (b) show double degeneracies.\ The horizontal panels next to (a) and (b) are the first three (solid, dash, and +) and four (with an extra $\times$) signal and idler mode probability densities in time.\ The labels of the insets in (a) and (b) are the same as the insets in Fig. \ref{fig2}, and we set in (a) $\delta p_{1,2}$ $=$ $\pm 60\Gamma_3$, $\delta p_3$ $=$ $0$, in (b) $\delta p_{1,2}$ $=$ $\delta p_{3,4}/3$ $=$ $\pm 30\Gamma_3$, and in both $\delta q_{m}$ $=$ $0$.\ The entropy of entanglement is denoted as S.}\label{fig3}
\end{figure}
%%%%%%%%%%%%%%%%%%%%%%%%%%%%%%%%%%%%%%%%%%%%%%%%%%%%%%%%%%%%%

We first investigate two frequency-multiplexed AE via Schmidt decomposition \cite{SM} in Fig. \ref{fig2} where the symmetric spectral function (to the line of $\Delta\omega_s$ $+$ $\Delta\omega_i$ $=$ $0$) is compared to the nonsymmetric one.\ In the Schmidt bases, we can express the biphoton state as $|\Psi\rangle$ $=$ $\sum_n\sqrt{\lambda_n}\hat{b}^\dag_n\hat{c}^\dag_n$ with eigenvalues $\lambda_n$, and signal $\hat b_n$ and idler $\hat c_n$ photon operators of mode functions $\psi_n(\omega_s)$ and $\phi_n(\omega_i)$ respectively.\ The spectral functions are set by $\delta p_2$ $=$ $-\delta p_1$ with $\delta q_{1,2}$ $=$ $0$, $2\delta p_{1,2}$, $\delta p_{1,2}$ for the left, right insets of (a), and the left inset of (d) respectively.\ The right inset of (d) is set by $\delta p_{1,2}$ $=$ $0$ with $\delta q_{2}$ $=$ $-\delta q_1$.\ Paired eigenvalues in the symmetric spectral function are characteristic of large entropy of entanglement S ($=$ $-$ $\sum_{n=1}^\infty$ $\lambda_n \rm{log}_2\lambda_n$) \cite{Bennett1996} while the nonsymmetric spectral function is more factorizable.\ The mode probability densities show degeneracies in frequency space as in (b) and (c) whereas the degeneracies break up when the frequency shifts are large enough that these two AE can be seen as independent \cite{SM}.\ For the nonsymmetric spectral function in (d), the mode probability density of the largest eigenvalue has a typical Gaussian for signal photon while a Lorentzian of two frequency peaks for the idler one.\ We note that not all the symmetric spectral functions have large S, for example of a combination of two insets of (a) in the setting of four multiplexed AE, which are more factorizable since they distribute symmetrically to the axes $\Delta\omega_{s,i}$ $=$ $0$.\ A factorizable biphoton state means the separability of the signal and idler modes.\ Therefore highly entangled biphoton states require the spectral functions to align along the energy-conserving axis $\Delta\omega_s$ $=$ $-\Delta\omega_i$.

Double degeneracies in frequency space indicate interference patterns which we demonstrate in Fig. \ref{fig3} for three and four AE with moderate frequency shifts.\ The eigenvalues of such symmetrical spectral function come in groups of $\rm N_{\rm MP}$ comparable values in which they form in pairs.\ As an example for four AE in (b), the mode probability densities of the same eigenvalues oscillate in the same period while they differ in phases.\ The oscillation period of the interference pattern provides a way to discriminate pairwise eigenfunctions.\ Furthermore we may even select one specific mode by a spectral filter if frequency shifts of our scheme are made large enough that $|\delta p_m|$ $\gtrsim$ $4/\tau$ and $4\Gamma_3^{\rm N}$.\ Signatures of Gaussian and Lorentzian are also obvious respectively in the signal and idler modes in time where idler modes have a long tail of oscillation determined by the Lorentzian linewidth.\ An increasing entropy of entanglement is expected when more spectral weights lie on the energy-conserving axis.

\begin{figure}[t]
\centering
\includegraphics[width=8.5cm,height=4.2cm]{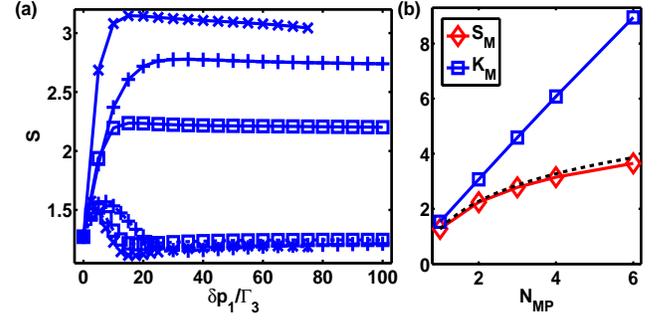}
\caption{(Color online) Entropy of entanglement (S) for N$_{\rm MP}$ $=$ $2$, $3$, $4$, and Schmidt number K.\ (a) We plot S as a function of $\delta p_1$ for symmetrical spectral functions for two to four AE (solid $\square$, +, and $\times$), indicating an increasing entanglement.\ For comparisons we set various $\delta p_m$ as equal separations with $\delta q_{m}$ $=$ $\delta p_{m}$ (dash $\square$, +, and $\times$) in the nonsymmetric spectra where S does not increase significantly as the number of AE grows.\ S for four AE (solid and dash $\times$) ends at $\delta p_1$ $=$ $75\Gamma_3$ before the spectral function goes out of the spectral range.\ (b) A linear growth of maximum Schmidt number K$_{\rm M}$ and logarithmic growth of maximum S, S$_{\rm M}$, compared to the dash line [S$_{\rm N_{\rm MP} =1}$ $+$ ${\rm log}_2({\rm N_{\rm MP}})$].}\label{fig4}
\end{figure}

For a comparison of entanglement growth in multiplexed AE, we plot S as a dependence of frequency shifts for two to four AE in Fig. \ref{fig4}.\ Nonsymmetric spectral functions show no significant increase of S as we multiplex more AE while for symmetric ones along the energy-conserving axis, the maximum entropy of entanglement S$_{\rm M}$ increases logarithmically in N$_{\rm MP}$.\ As a comparison we plot Schmidt number K $=$ $1/\sum_n\lambda_n^2$ \cite{Grobe1994, Law2004} versus N$_{\rm MP}$, which indicates an averaging measure of biphoton correlation that scales linearly as we multiplex more numbers of AE.
%%%%%%%%%%%%%%%%%%%%%%%%%%%%%%%%%%%%%%%%%%%%%%%%%%%%%%%%%%%%%
%%%%%%%%%%%%%%%%%%%%%%%%%%%%%%%%%%%%%%%%%%%%%%%%%%%%%%%%%%%%%

To describe the trend of entropy of entanglement growth as N$_{\rm MP}$, we may approximately analyze the eigenvalues in the following.\ We observe that the multiplexed AE has eigenvalues in pairs where first N$_{\rm MP}$ eigenvalues are comparable.\ We may make a cutoff to N$_{\rm MP}$ eigenvalues and assume they are equal and renormalized.\ In this way we end up with a maximally entangled state in discrete bases, regardless of the continuous frequency entanglement, which reads
\begin{eqnarray}
|\Psi\rangle_{\rm d} = \sum_{m=1}^{\rm N_{\rm MP}}\frac{1}{\sqrt{{\rm N_{\rm MP}}}}\hat{a}_{s,m}^\dag\hat{a}_{i,m}^\dag|0\rangle,
\end{eqnarray}
where signal and idler modes denoted by m also terminate at the cutoff eigenvalues.\ The entropy of entanglement of this sort is easily calculated as S$_{\rm d}$ $=$ $\rm log_2({\rm N_{\rm MP}})$.\ To account for the continuous frequency entanglement, we add to S$_{\rm d}$ with an excess entropy S$_{\rm ex}$ $=$ S$_{\rm N_{\rm MP} =1}$ which is the entropy of entanglement for each individual atomic ensemble in continuous frequency space.\ Compare with the Hilbert space of the maximally entangled qudit state of dimensions $\rm N_{\rm MP}$, which is $2^{\rm N_{\rm MP}}$, the excess entropy provides an extra scaling enhancement of $2^{\rm S_{\rm ex}}$ to the discrete system.\ In Fig. \ref{fig4}(b) we show that the maximum entropy of entanglement in the multiplexed scheme with symmetric spectral functions approximately follows the curve of S$_{\rm d}$ $+$ S$_{\rm ex}$.\ Therefore for our scheme of multiplexed cold AE, the entropy can increase either by multiplexing more AE or generating a more spectrally entangled biphoton source.\ The excess entropy can be increased by increasing the superradiant decay rate $\Gamma_3^{\rm N}$, equivalently the optical density of AE, or using shorter driving fields (smaller $\tau$) \cite{Jen2012-2}.

The scheme of multiplexed AE by frequency shifters allows for frequency encoding/decoding, applicable in quantum key distribution \cite{Gisin2002} in continuous degrees of freedom.\ We may address the coding individually for just signal, idler, or collaboratively for both modes in nonsymmetric or symmetric spectral functions.\ According to Fig. \ref{fig4}, the entropy of entanglement saturates as frequency shifts increase such that the first N$_{\rm MP}$ modes are well-separated in frequency domains.\ We then can activate the coding spectrally while keeping the entropy of entanglement intact.\ Potentially the biphoton state of the multiplexed scheme can be even manipulated in their amplitude and phase information in designated frequency bins \cite{Bernhard2013}, giving a coherent control of high-dimensional continuous frequency entanglement.\ Similar to Hadamard codes carried out in entangled photons of spontaneous parametric down conversion in frequency \cite{Lukens2014} or equivalently in time domains \cite{Belthangady2010}, our multiplexed cascade emissions can be in principle implement the codes by encoding their sequences in the signal photon and decoding in the idler one via coincidence measurements.\ 

%%%%%%%%%%%%%%%%%%%%%%%%%%%%%%%%%%%%%%%%%%%%%%%%%%%%%%%%%%%%%%%%%%%%%%%%%%%%%%%%%%%%%%%%%%%%%%%%%%%%%%%%%%%%%%%%%%%%

In conclusion, we propose a scheme of multiplexed cascade emissions from cold atomic ensembles via frequency shifters, in which we may shape its spectral properties and take advantage of its telecom bandwidth useful for long-distance quantum communication.\ The Schmidt decomposition of the biphoton state shows pairwise eigenvalues that may have interference patterns with moderate frequency shifts.\ The entropy of entanglement of such scheme can be larger if the spectral function of the biphoton state is manipulated along the energy-conserving axis or if more atomic ensembles are multiplexed.\ This multiplexed biphoton scheme provides a flexible control over continuous spectral entanglement and offers a potentially robust candidate of long-distance quantum communication applicable for spectral coding and quantum key distribution.\ Since the performance of our scheme heavily depend on the number of multiplexed atomic ensembles, we may also make use of optical lattices and their controllability, which are useful for a large-scale implementation.

%{\it Acknowledgments.--} 
We gratefully acknowledge funding by the Ministry of Science and Technology, Taiwan, under Grant No. MOST-101-2112-M-001-021-MY3, and the support of NCTS on this work.\ We are also grateful for fruitful discussions with S.-Y. Lan.
%%%%%%%%%%%%%%%%%%%%%%%%%%%%%%%%%%%%%%%%%%%%%%%%%%%%%%%%%%%%%%%%%%%%%%%%%%%%%%%%%%%%%%%%%%%%%%%%%%%%%%%%%%%%%%%%%%%%

%%%%%%%%%%%%%%%%%%%%%%%%%%%%%%%%%%%%%%%%%%%%%%%%%%%%%%%%%%%%%%%%%%%%%%%%%%%%%%%%%%%%%%%%%%%%%%%%%%%%%%%%%%%%%%%%%%%%
\clearpage
\section*{Supplemental Materials for Spectral Shaping of Cascade Emissions from Multiplexed Cold Atomic Ensembles}

\section{Schr\"odinger equation of motion}
From the Hamiltonian $V_{\rm I}(t)$ in the main paper, we assume large detuned and weak driving fields satisfying $\Delta_1\gg\sqrt{N}|\Omega_a|$.\ Therefore only single excitation is considered, and we ignore spontaneous decay during excitations.\ The state function can be described by \cite{Jen2012-2}
\begin{eqnarray}
|\psi(t)\rangle&=&\mathcal{E}(t)|0,{\rm vac}\rangle+\sum^N_{\mu=1} A_\mu(t)|1_\mu,vac\rangle\nonumber\\
&&+\sum^N_{\mu=1} B_\mu(t)|2_\mu,vac\rangle+\sum^N_{\mu=1}\sum_{s}C^\mu_s(t)|3_\mu,1_{\k_s,\lambda_s}\rangle\nonumber\\
&&+\sum_{s,i}D_{s,i}(t)|0,1_{\k_s,\lambda_s},1_{\k_i,\lambda_i}\rangle,
\end{eqnarray}
where $s(i)$ indexed in the summation denotes $k_{s(i)}$, $\lambda_{s(i)}$ respectively, $|m_\mu\rangle$ $\equiv$ $|m_\mu\rangle|0\rangle^{\otimes N-1}_{\nu\neq\mu}$ with $m$ $=$ $1$, $2$, $3$, and $|{\rm vac}\rangle$ is the vacuum photon state.\ The probability amplitudes are $\mathcal{E}(t)$, $A_{\mu}(t)$, $B_{\mu}(t)$, $C_{s}^{\mu}(t)$,  $D_{s,i}(t)$, which indicate the complete cycle of single excitation process from the ground state to intermediate and upper excited states, then intermediate excited state with emission of a signal photon, and the ground state with signal and idler emissions.\ We formulate Schr\"odinger equation of motion by applying $i\hbar\frac{\partial}{\partial t}|\psi(t)\rangle$ $=$ $V_{\rm I}(t)|\psi(t)\rangle$, and we have
\begin{eqnarray}
i\dot{\mathcal{E}}&=&-\frac{\Omega_a^*}{2} \sum_\mu e^{-i\k_a\cdot\r_\mu}A_\mu\label{eqn1},\\
i\dot{A}_\mu&=&-\frac{\Omega_a}{2}e^{i\k_a\cdot\r_\mu}\mathcal{E}-\frac{\Omega_b^*}{2} e^{-i\k_b\cdot\r_\mu}B_\mu-\Delta_1 A_\mu\label{eqn2},\\
i\dot{B}_\mu&=&-\frac{\Omega_b}{2}e^{i\k_b\cdot\r_\mu}A_\mu-\Delta_2 B_\mu-i\sum_{\k_s,\lambda_s}g_{s}e^{i\k_s\cdot\r_\mu}\nonumber\\
&&\times e^{-i\Delta\omega_{s}t}C^\mu_{s},\label{eqn3}\\
\dot{C}^\mu_{s}&=&ig_{s}^*e^{-i\k_s\cdot\r_\mu}e^{i\Delta\omega_{s}t}B_\mu-i\sum_{\k_i,\lambda_i}g_{i}e^{i\k_i\cdot\r_\mu}\nonumber\\
&&\times e^{-i\Delta\omega_{i}t}D_{s,i},\label{eqn4}\\
i\dot{D}_{s,i}&=&ig_{i}^*\sum_\mu e^{-i\k_i\cdot\r_\mu}e^{i\Delta\omega_{i}t}C^\mu_{s}.\label{eqn5}
\end{eqnarray}

In the limit of large detunings, $|\Delta_{1,2}|$ $\gg$ $|\Omega_{a,b}|$, $\Gamma_{2,3}$, where $\Gamma_{2,3}$ is the intrinsic decay rate for the upper and intermediate excited states respectively, we can solve the coupled equations of motion by adiabatically eliminating the intermediate and upper excited states.\ The adiabatic approximation \cite{QO:Scully} requires slowly varying driving pulses, which is equally to solve for the steady state solutions of the above coupled equations perturbatively.\ The ground state probability is derived to be unity in the zeroth order of perturbation, and other probability amplitudes are
\begin{eqnarray}
A_\mu(t)&\approx& -\frac{\Omega_a(t)}{2\Delta_1}e^{i\k_a\cdot\r_\mu},\\
B_\mu(t)&\approx& \frac{\Omega_a(t)\Omega_b(t)}{4\Delta_1\Delta_2}e^{i(\k_a+\k_b)\cdot\r_\mu},
\end{eqnarray}
where the singly-excited atom follows the driving fields to the intermediate and the upper excited states.

Substituting Eq. (\ref{eqn5}) into Eq. (\ref{eqn4}), we have
\begin{eqnarray}
\dot{C}^\mu_{s}(t)&=&g_s^*e^{-i\k_s\cdot\r_\mu}e^{i\Delta\omega_{s}t}B_\mu(t)-\sum_\nu\sum_{\k_i,\lambda_i}|g_i|^2e^{i\k_i\cdot(\r_\mu-\r_\nu)}\nonumber\\
&&\times\int_{-\infty}^t dt'e^{i\Delta\omega_{i}(t'-t)}C^\nu_{s}(t').
\end{eqnarray}
We now consider a symmetric state basis that contributes to the biphoton state generation most significantly, and define a phased probability amplitude $C_{s,\k_{i}}$ $=$ $\sum_{\mu}C_{s}^{\mu}e^{-i\k_{i}\cdot\r_{\mu}}$, which becomes
\begin{eqnarray}
C_{s,k_{i}}(t)&=&g_{s}^{\ast}\sum_{\mu}e^{i\Delta\k\cdot\r_{\mu}}\int_{-\infty}^{t}dt^{\prime}e^{i\Delta\omega_{s}t^{\prime}}\nonumber\\
&&\times e^{(-\frac{\Gamma_{3}^{\rm N}}{2}+i\delta\omega_{i})(t-t^{\prime})}b(t^{\prime}),
\end{eqnarray}
where $\Gamma_{3}^{\rm N}$ $=$ $(\rm{N}\bar{\mu}+1)\Gamma_{3}$ is a superradiant decay rate for the atomic transition $|3\rangle$ $\rightarrow$ $|0\rangle$.\ The geometrical constant $\bar{\mu}$ \cite{Lehmberg1970} depends on the shape of the atomic ensemble, and the cooperative Lamb shift (CLS) \cite{Friedberg1973} is denoted as $\delta\omega_{i}$,
\begin{eqnarray}
\delta\omega_{i}&\equiv&\int_{0}^{\infty}d\omega\frac{\Gamma}{2\pi}\Big[\textrm{P.V.}(\omega-\omega_{3})^{-1}\Big]N\bar{\mu}(\k),
\end{eqnarray}
where P.V. is principal value, and $\Gamma$ $=$ $|d|^2\omega^3/(3\pi\hbar\epsilon_0 c^3)$ with a dipole moment $d$.\ Note that here we have renormalized the Lamb shift of the transition, and a complete formulation of the CLS requires non-RWA (rotating-wave approximation) terms in the Hamiltonian, which contributes to $\delta\omega_i$ with an additional term proportional to P.V.$(\omega_i+\omega_3)^{-1}$.\ The CLS in the above has an integral dependence of a spontaneous decay rate $\propto$ $\rm{N}\bar{\mu}(\k)$, which is a Hilbert transform \cite{Cohen-Tannoudji1992} if the Lamb shift is put back into $\delta\omega_i$. 

Finally we have the biphoton state probability amplitude,
\begin{eqnarray}
D_{s,i}(t)&=& g_{i}^{\ast}g_{s}^{\ast}\sum_{\mu}e^{i\Delta\k\cdot\r_{\mu}}\int_{-\infty}^{t}\int_{-\infty}^{t^{\prime}}dt^{\prime\prime}dt^{\prime}
e^{i\Delta\omega_{i}t^{\prime}}e^{i\Delta\omega_{s}t^{\prime\prime}}\nonumber\\
&&\times b(t^{\prime\prime})e^{(-\frac{\Gamma_{3}^{\rm N}}{2}+i\delta\omega_{i})(t^{\prime}-t^{\prime\prime})},\label{Dsi}
\end{eqnarray}
where $b(t)=\frac{\Omega_{a}(t)\Omega_{b}(t)}{4\Delta_{1}\Delta_{2}}$ is proportional to the product of the Rabi frequencies.\ The factor $\sum_{\mu}e^{i\Delta\k\cdot\r_{\mu}}$ reflects the phase-matching condition of four-wave mixing when
the wavevector mismatch $\Delta\k$ $=$ $\k_{a}$ $+$ $\k_{b}$ $-$ $\k_{s}$ $-$ $\k_{i}$ $\rightarrow$ $0$. 

We consider normalized Gaussian pulses where $\Omega_{a}(t)$ $=$ $\frac{1}{\sqrt{\pi}\tau}\tilde{\Omega}_{a}e^{-t^{2}/\tau^{2}}$, $\Omega_{b}(t)$ $=$ $\frac{1}{\sqrt{\pi}\tau}\tilde{\Omega}_{b}e^{-t^{2}/\tau^{2}}$ with the same pulse width. $\tilde{\Omega}_{a,b}$ is the pulse area.\ In the long time limit, we have the probability amplitude $D_{si}$ after the integration of Eq. (\ref{Dsi}),
\begin{eqnarray}
D_{si}(\Delta\omega_{s},\Delta\omega_{i})&=&\frac{\tilde{\Omega}_{a}\tilde{\Omega}_{b}g_{i}^{\ast}g_{s}^{\ast}}{4\Delta_{1}%
\Delta_{2}}\frac{\sum_{\mu}e^{i\Delta\k\cdot\r_{\mu}}}{\sqrt{2\pi}\tau}\nonumber\\
&&\times\frac{e^{-(\Delta\omega_{s}+\Delta\omega_{i})^{2}\tau^{2}/8}}{\frac{\Gamma_{3}^{N}}{2}-i\Delta\omega_{i}},
\end{eqnarray}
which indicates a spectral width $\Gamma_{3}^{N}/2$ for a Lorentzian idler photon modulating a Gaussian profile with a spectral width
$2\sqrt{2}/\tau$ for signal and idler photons.\ The maximum condition in Gaussian profile of $\Delta\omega_{s}$ $+$ $\Delta\omega_{i}$ $=$ $0$ means the energy conservation of signal and idler photons with two driving fields at their central frequencies.\ The CLS is negligible (in the order of kHz) \cite{Jen2015} in general for our spectral shaping (frequency shifts of MHz) in conventional cold atomic ensembles (AE).
%%%%%%%%%%%%%%%%%%%%%%%%%%%%%%%%%%%%%%%%%%%%%%%%%%%%%%%%%%%%%%%%%%%%%%%%%%%%%%%%%%%%%
\section{Schmidt decomposition}
The multimode analysis and entanglement properties of our multiplexed cascade emissions can be done by Schmidt decomposition.\ Here we review the theoretical background of Schmidt decomposition in frequency space.\ For some specific polarizations $\lambda_{s}$ and $\lambda_{i}$, we have the biphoton state vector $|\Psi\rangle$ with a spectral function $f(\omega_{s},\omega_{i})$,
\begin{equation}
|\Psi\rangle=\int f(\omega_{s},\omega_{i})\hat{a}_{\lambda_{s}}^{\dag}(\omega_{s})\hat{a}_{\lambda_{i}}^{\dag}(\omega_{i})|0\rangle d\omega
_{s}d\omega_{i}.
\end{equation}

Following the theoretical work on two-photon pulses generated from parametric down-conversion by Law {\it et al.} \cite{law}, the quantification of entanglement can be determined in the Schmidt basis where the state vector is expressed as
\begin{eqnarray}
|\Psi\rangle&=&\sum_{n}\sqrt{\lambda_{n}}\hat{b}_{n}^{\dag}\hat{c}_{n}^{\dag}|0\rangle,\\
\hat{b}_{n}^{\dag}&\equiv&\int\psi_{n}(\omega_{s})\hat{a}_{\lambda_{s}}^{\dag}(\omega_{s})d\omega_{s},\\
\hat{c}_{n}^{\dag}&\equiv&\int\phi_{n}(\omega_{i})\hat{a}_{\lambda_{i}}^{\dag}(\omega_{i})d\omega_{i},
\end{eqnarray}
where $\hat{b}_{n}^{\dag},$ $\hat{c}_{n}^{\dag}$ are effective creation operators and $\lambda_n$'s (no confusion with polarization index $\lambda_{s,i}$) are probabilities in corresponding biphoton mode $n$.\ Eigenvalues $\lambda_{n}$, and eigenfunctions $\psi_{n}$, $\phi_{n},$ are the solutions of the eigenvalue equations,
\begin{eqnarray}
\int K_{1}(\omega,\omega^{\prime})\psi_{n}(\omega^{\prime})d\omega^{\prime}  &=&\lambda_{n}\psi_{n}(\omega),\\
\int K_{2}(\omega,\omega^{\prime})\phi_{n}(\omega^{\prime})d\omega^{\prime}  &=&\lambda_{n}\phi_{n}(\omega),
\end{eqnarray}
where 
\begin{eqnarray}
K_{1}(\omega,\omega^{\prime}) &\equiv&\int f(\omega,\omega_{1})f^{\ast}(\omega^{\prime},\omega_{1})d\omega_{1}, \\
K_{2}(\omega,\omega^{\prime}) &\equiv&\int f(\omega_{2},\omega)f^{\ast}(\omega_{2},\omega^{\prime})d\omega_{2}, 
\end{eqnarray}
which are the kernels for the one-photon spectral correlations \cite{law,parker}.\ Orthogonality of eigenfunctions is assured that  $\int\psi_{i}(\omega)$$\psi_{j}(\omega)d\omega$ $=$ $\delta_{ij}$, $\int\phi_{i}(\omega)$$\phi_{j}(\omega)d\omega$ $=$ $\delta_{ij},$ and the normalization of quantum state requires $\sum_{n}\lambda_{n}$ $=$ $1$.

In the Schmidt basis, the entanglement entropy can be expressed as
\begin{equation}
S=-\sum_{n=1}^{\infty}\lambda_{n}\textrm{log}_2\lambda_{n}.
\end{equation}
If there is only one non-zero Schmidt number that $\lambda_{1}=1$, the entropy is zero, which means no entanglement, or a factorizable state. \ For more than
one non-zero Schmidt numbers, the entropy is larger than zero, and the bipartite entanglement is nonvanishing.\ Finite entanglement means $f(\omega_{s},\omega_{i})$ cannot be factorized as $g(\omega_{s})h(\omega_{i}),$ a multiplication of two separate spectral functions.
%%%%%%%%%%%%%%%%%%%%%%%%%%%%%%%%%%%%%%%%%%%%%%%%%%%%%%%%%%%%%%%%%%%%%%%%%%%%%%%%%%%%%%%%%%%%%%%%%%%%%%%%%%%%%%%%%%%%
\section{Schmidt decomposition for two multiplexed atomic ensembles}
%%%%%%%%%%%%%%%%%%%%%%%%%%%%%%%%%%%%%%%%%%%%%%%%%%%%%%%%%%%%%
\begin{figure}[b]
\centering
\includegraphics[width=8.5cm,height=4.5cm]{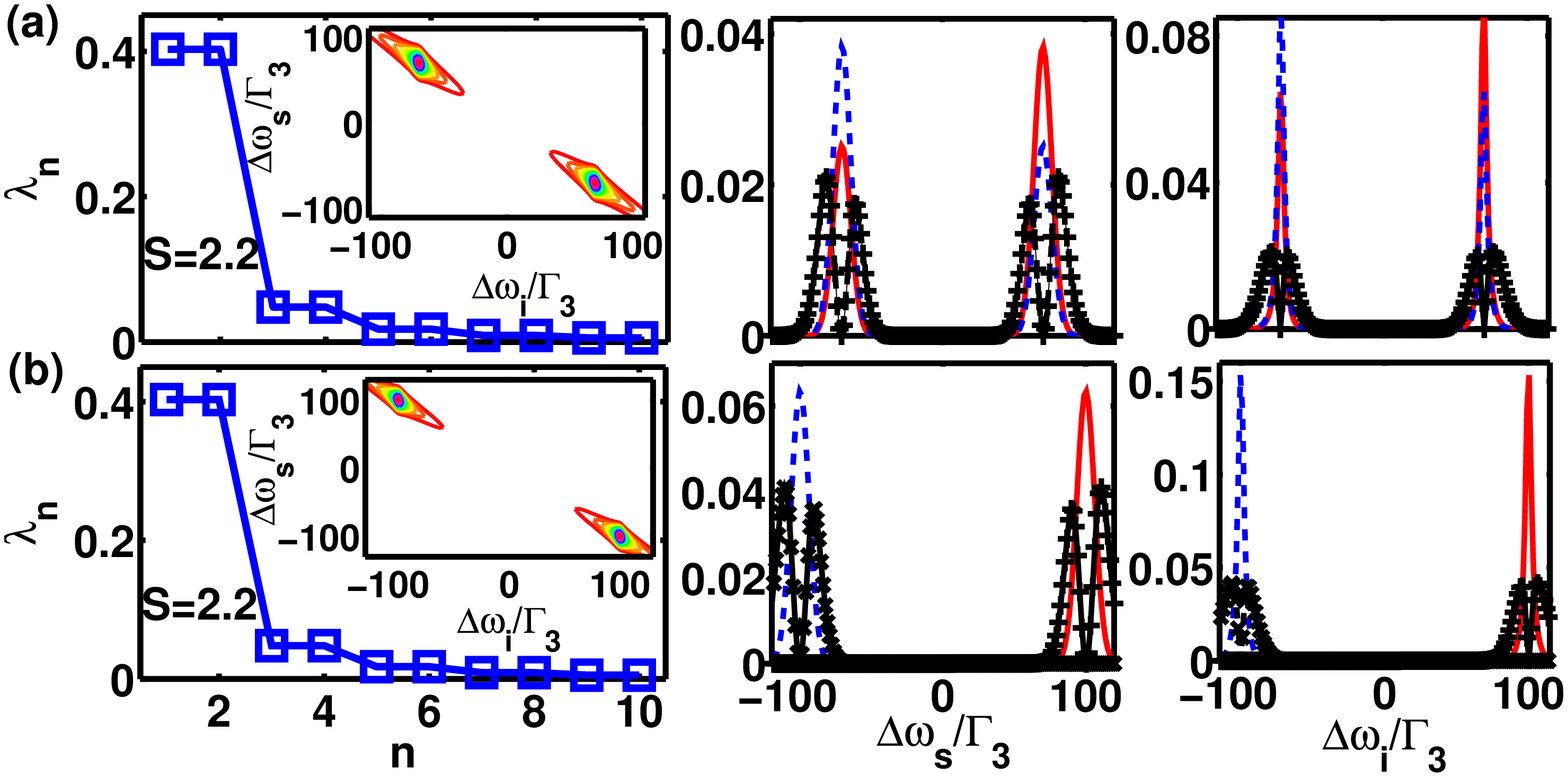}
\caption{(Color online) Multimode biphoton mode probability densities $|\psi_n(\omega_s)|^2$, $|\phi_n(\omega_i)|^2$, and entanglement entropy for two AE in the symmetric spectral functions.\ We set $\delta p_1$ $=$ $-$ $\delta p_2$ and $\delta q_{1,2}$ $=$ $0$ for (a) $\delta p_1$ $=$ $70\Gamma_3$ and (b) $\delta p_1$ $=$ $100\Gamma_3$.\ Four mode probability densities are plotted correspondingly (solid, dash, +, and $\times$) while the third and fourth ones in (a) are degenerate (only ``+" is marked).}\label{s1}
\end{figure}
%%%%%%%%%%%%%%%%%%%%%%%%%%%%%%%%%%%%%%%%%%%%%%%%%%%%%%%%%%%%%
Here we demonstrate how the spectral shaping changes the mode probability densities as the phase modulations increase for two AE.\ In Fig. \ref{s1}, we consider the symmetric spectral functions.\ As the phase modulation increases from Fig. \ref{s1}(a) to (b), we see the first four mode probability densities start to separate into frequency-resolved waveforms.\ In addition, the discrimination process starts from the first two modes as in Fig. \ref{s1}(a) where they are partially overlapped while the third and fourth modes are still not able to be distinguished from each other.\ Note that the Schmidt numbers show in pairs with double degeneracies but the signal and idler mode probability densities are only degenerate before the frequency shifts are too large.

\begin{figure}[t]
\centering
\includegraphics[width=8.5cm,height=4.5cm]{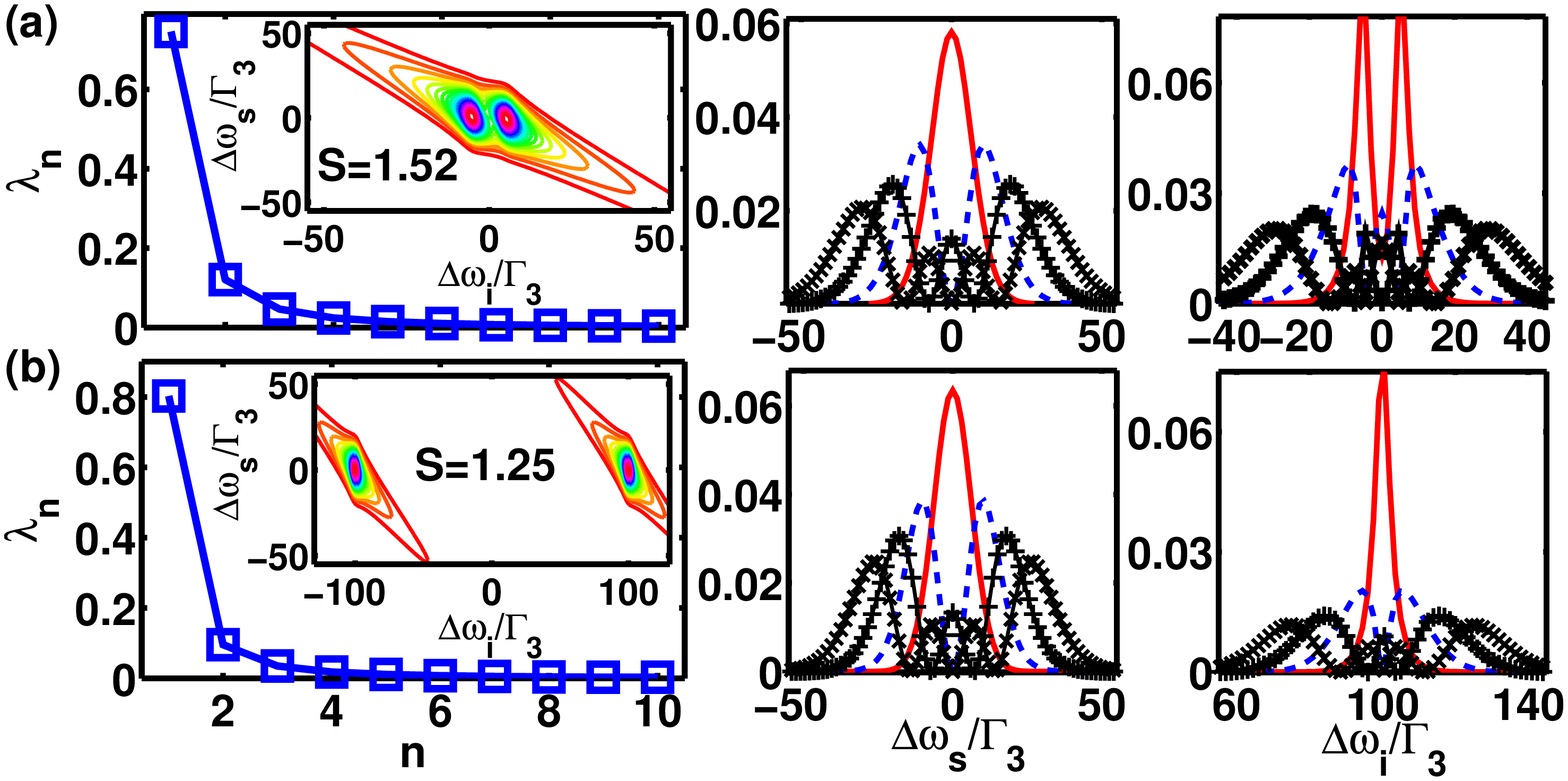}
\caption{(Color online) Multimode biphoton mode probability densities and entanglement entropy for two AE in the nonsymmetric spectral functions.\ We set $\delta p_1$ $=$ $-$ $\delta p_2$ and $\delta q_{1,2}$ $=$ $\delta p_{1,2}$ for (a) $\delta p_1$ $=$ $10\Gamma_3$ and (b) $\delta p_1$ $=$ $100\Gamma_3$.\ Four mode probability densities are plotted correspondingly (solid, dash, +, and $\times$) while the idler mode probability density in (b) is demonstrated only for the positive frequency side which is symmetric to its negative counterpart.}\label{s2}
\end{figure}

In Fig. \ref{s2}, we consider the nonsymmetric spectral functions on the line of $\Delta\omega_s$ $=$ $0$.\ The mode probability densities are not degenerate with each other, which also reflects in the Schmidt number distributions.\ As the phase modulation increases, the double peaks of the idler modes appear to separate in frequency domains as expected.\ We note that idler modes are symmetric to $\Delta\omega_i$ $=$ $0$.
%%%%%%%%%%%%%%%%%%%%%%%%%%%%%%%%%%%%%%%%%%%%%%%%%%%%%%%%%%%%%
%%%%%%%%%%%%%%%%%%%%%%%%%%%%%%%%%%%%%%%%%%%%%%%%%%%%%%%%%%%%%
%%%%%%%%%%%%%%%%%%%%%%%%%%%%%%%%%%%%%%%%%%%%%%%%%%%%%%%%%%%%%%%%%%%%%%%%%%%%%%%%%%%%%%%%%%%%%%%%%%%%%%%%%%%%%%%%%%%%

\end{document}